\documentclass[12pt]{article}

\usepackage{ol2}
\usepackage{amsmath}

\begin{document}

%\twocolumn[
\title{Demonstration of coherent emission from high-$\beta$ photonic crystal nanolasers at room temperature}

\author{R. Hostein, R. Braive, L. Le Gratiet, A. Talneau, G. Beaudoin, I. Robert-Philip, I. Sagnes and A. Beveratos}

\address{
Laboratoire de Photonique et Nanostructures LPN-CNRS UPR-20, Route de Nozay, 91460 Marcoussis, France\\
$^*$Corresponding author: Alexios.Beveratos@lpn.cnrs.fr }
%\email{Alexios.Beveratos@lpn.cnrs.fr}

\begin{abstract}

We report on lasing at room temperature and at telecommunications wavelength from photonic crystal nanocavities based on InAsP/InP quantum dots. Such laser cavities with a small modal volume and high quality factor display a high spontaneous emission coupling factor ($\beta$). Lasing is confirmed by measuring the second order autocorrelation function. A smooth transition from chaotic to coherent emission is observed, and coherent emission is obtained at 8 times the threshold power.
\end{abstract}

 %]

\maketitle

The continuous down scaling of the design and fabrication of microcavities\cite{Song2005} has opened the road towards the realization of integrated laser sources, with a small footprint and low power consumption, for optical interconnects, or lab on-chip applications \cite{MOMA}. When the cavity length is of the order of the emission wavelength, cavity quantum electrodynamic effects (cQED) cannot not be neglected anymore, introducing a modification of the static \cite{Yamamoto1991} and dynamic \cite{Atlug2006,Braive2009} properties of such devices. In the weak coupling regime, the acceleration and spatial redistribution of spontaneous emission lead to the engineering of non-conventional lasers with high spontaneous emission coupling factor $\beta$. For such laser structures, the classical definition of threshold is expected to be ill-defined and the transition from spontaneous to stimulated emission extends over a wide excitation power range \cite{Yamamoto1991}. Onset of lasing can be inferred from the narrowing of the emission linewidth, following the Shallow-Townes equations as shown by \cite{Ates2008} at 4 K and \cite{Nomura2007} at room temperature with InAs/GaAs quantum dots as gain material. But in both cases due to the important $\beta$ factor, the linewidth was only slightly reduced which did not allow to deduce unambiguously onset of a coherent emission. The sole unambiguous signature of lasing in such high-$\beta$ lasers can only be deduced from the statistics of the emitted light with a transition from chaotic to coherent emission. Such demonstration has been achieved on systems operating at 4 K \cite{Strauf2006, Ulrich2007, Choi2007}. Considerable efforts toward the realization of nanocavity based lasers at room temperature have been undertaken, using quantum wells under pulsed \cite{Lu2009}, or continuous wave excitation \cite{Nozaki2007,Watanabe2008, Vecchi2007,Nomura2007}. Recently pulsed \cite{Bordas2009} and continuous wave operation \cite{Martinez2009} has been reported with with InAs/InP quantum dots and quantum wires as gain medium respectively. In all the above mentioned realizations, lasing has been inferred by observing a non-linear increase in the number of emitted photons during the transition which is not sufficient  to deduce the emission of coherent light. In this paper we demonstrate, by measuring the second order autocorrelation function, monomode laser emission at room temperature from InAsP/InP quantum dot nanolaser, subjected to cQED effects, under pulsed and continuous wave excitation. We observe a 30 dB signal to noise ratio, and a smooth transition from chaotic to coherent emission. It is noteworthy that coherent laser emission is only obtained at a pump power higher than eight times the threshold power defined by the kink in the Light-In Light-Out curve (L-L).

The laser cavity is formed by a photonic crystal double
heterostructure \cite{Song2005} etched on a 320 nm-thick suspended
InP membrane \cite{Hostein2009} and incorporating a single layer of self-assembled
InAsP quantum dots \cite{Michon2008} at its vertical center plane. The whole
structure is grown by metalo-organic chemical vapor deposition. The quantum dot
density is of the order of 15.10$^{9}$ cm$^{-2}$ and their
spontaneous emission is centered around 1560 nm at 300 K with an
inhomogeneous broadening of about 150 nm. The cavity is fabricated
using electron beam lithography, inductively coupled plasma
etching and wet etching \cite{Talneau2008}. The structure consists of a W1 waveguide composed of one
missing row of holes in the $\Gamma$-K direction of a hexagonal
lattice structure, with a local enhancement in the lattice period
over two periods at the center of the photonic crystal
waveguide. The lattice period is only modified along the $\Gamma$-K
direction. The waveguide with the larger longitudinal lattice
constant a$_l$=440 nm forms the nanocavity closed by two
surrounding mirror waveguides with smaller lattice constant
a$_m$=410 nm. The target air hole-radius r /a$_m$ is 0.293. The cold cavity quality factor is measured to be Q=45000 \cite{Hostein2009}, with an effective volume of 1.3$(\lambda/n)^3$ measured by FDTD from an scanning electron microscope image.

The samples are studied at room temperature. Optical excitation is
provided by either a continuous wave (CW) 532 nm Nd:YAG laser or a
pulsed Ti:Sa laser emitting at 840 nm with a repetition rate of 80
MHz and a 5 ps pulse width. The pumping laser is focused to a 5
$\mu$m spot on the sample by a microscope objective (Numerical
Aperture = 0.4). Taking in account the Fresnel reflexion on the InP interface (R=30\% at normal incidence), and the effective absorption of InP (12900 cm$^{-1}$), the reflexion on the InP substrate and the filling factor of the photonic crystal, only 22\% of the impimping light is actually absorbed. The luminescence is collected by the same microscope objective and separated from the pumping laser by means of a dichroic mirror and an antireflection coated silicon filter. The spontaneous emission is spectrally dispersed by a 0.5 m spectrometer with a spectral resolution of 0.15 nm and detected by
a cooled InGaAs photodiodes array (Roper Scientific). The second order autocorrelation function is obtained by a standard Hanbury-Brown and Twiss setup, using two superconducting single photon counters (SSPD-Scontel). Each detector has a time
resolution of $\sqrt{2}\sigma=70$ ps, a quantum efficiency of 3\% at 1.55 $\mu$m and dark count rates lower than 30 counts.s$^{-1}$. The histogram of the time intervals between two subsequent photons, one in each SSPD, is recorded on a Lecroy 8620A oscilloscope. Before being detected by the superconducting detector, the emission light is filtered by a tunable filter
(Santec, 0.4 nm bandwidth, 20 dB extinction rate).

\begin{figure}[h!]
   \begin{center}
   \begin{tabular}{c}
   \includegraphics[width=8.3cm]{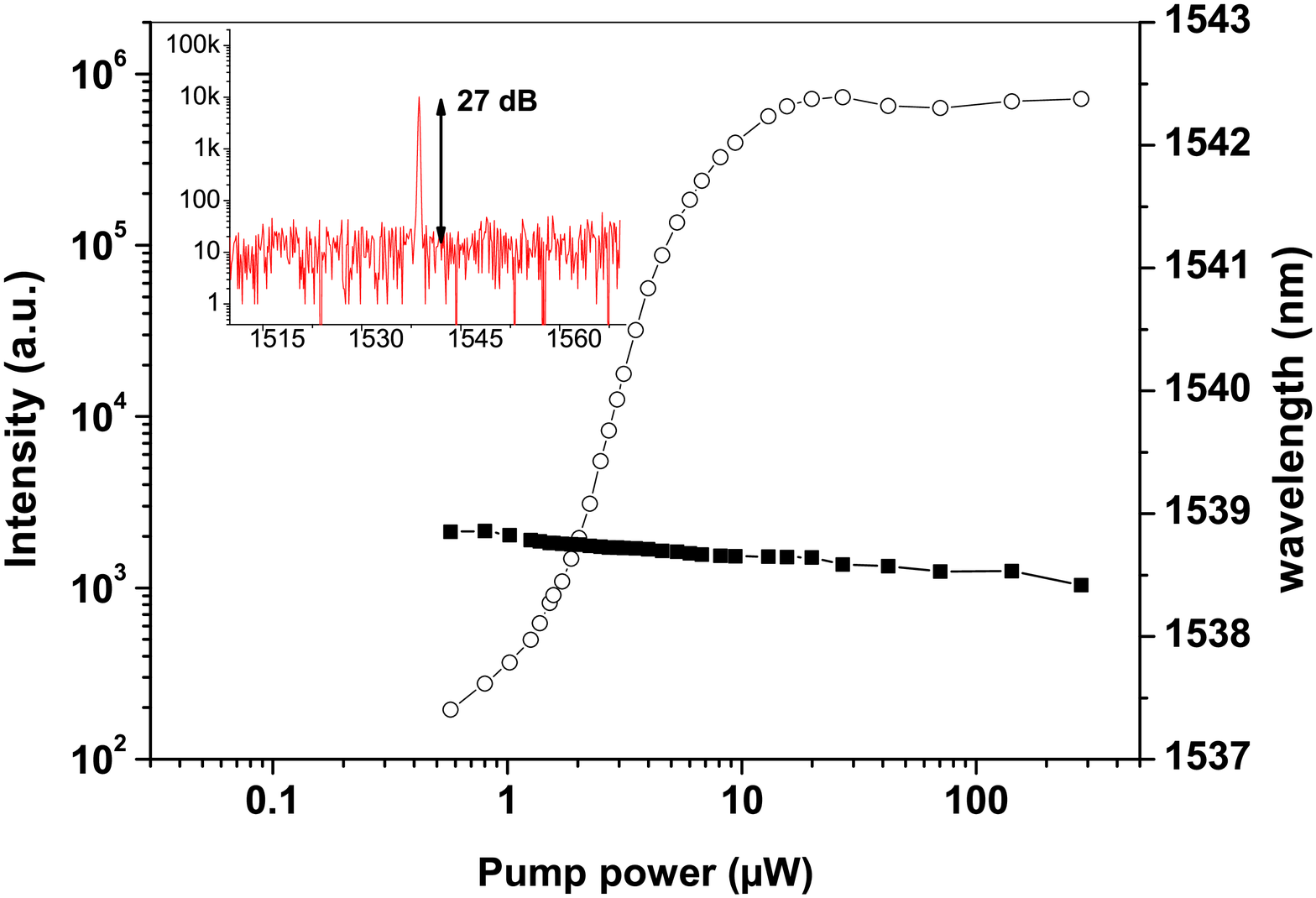}
   \end{tabular}
   \end{center}
   \caption[example]
   { \label{Fig1} (Left) Light-In Light-Out curve under pulsed excitation. (Right) Emission wavelength as function of excitation power. (Inset) Emission spectrum under an excitation power of 10$\mu$W.}
   \end{figure}

\begin{figure}[h!]
   \begin{center}
   \begin{tabular}{c}
   \includegraphics[width=8.3cm]{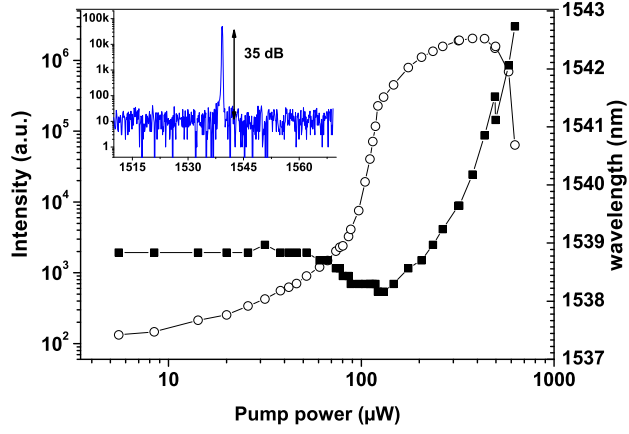}
   \end{tabular}
   \end{center}
   \caption[example]
   { \label{Fig2} (Left) Light-In Light-Out curve under CW excitation. (Right) Emission wavelength. (Inset) Emission spectrum under an excitation power of 300$\mu$W}
   \end{figure}

Figures \ref{Fig1} and \ref{Fig2} represent the standard light-in light-out curves (L-L) under pulsed and continuous wave excitation respectively. The insets of either figure show a clear monomode emission with a sideband rejection of 27 and 35 dB respectively. We will consider that the threshold power corresponds to the kink of the L-L curve. The effective threshold power is P$^{Pulse}_{th}$=3.15 $\mu$W (ie 39 fJ) under pulsed excitation, and P$^{CW}_{th}$=115 $\mu$W under continuous wave excitation. The transition from spontaneous to stimulated emission is relatively large in both cases and extends over \{0.38;0.40\}*Pth to \{2.68;1.77\}*Pth for \{pulsed;CW\} operation, indicating that the laser operates in the high-$\beta$ regime \cite{Nozaki2007}. Under pulsed excitation the emission wavelength shows an linear, in log x-scale, decrease as a function of excitation power, due to the increasing carrier density. Conversely, under continuous wave excitation, the emission wavelength shows a blueshift for low excitation power. Once the laser regime is reached, the emission wavelengths shifts towards the longer wavelength, due to important thermal effects. Simultaneously, an increase of the emission linewidth is observed (not represented) most probably due to important thermal fluctuations. Thermal degradation of the laser emission is observed at a pump power of P$^{CW}_{max}=$435 $\mu$W under CW operation, while under pulsed excitation the laser emission saturates due to gain material saturation, and is sustained even for pump powers two orders of magnitude above the threshold value.

\begin{figure}[h!]
   \begin{center}
   \begin{tabular}{c}
   \includegraphics[width=8.3cm]{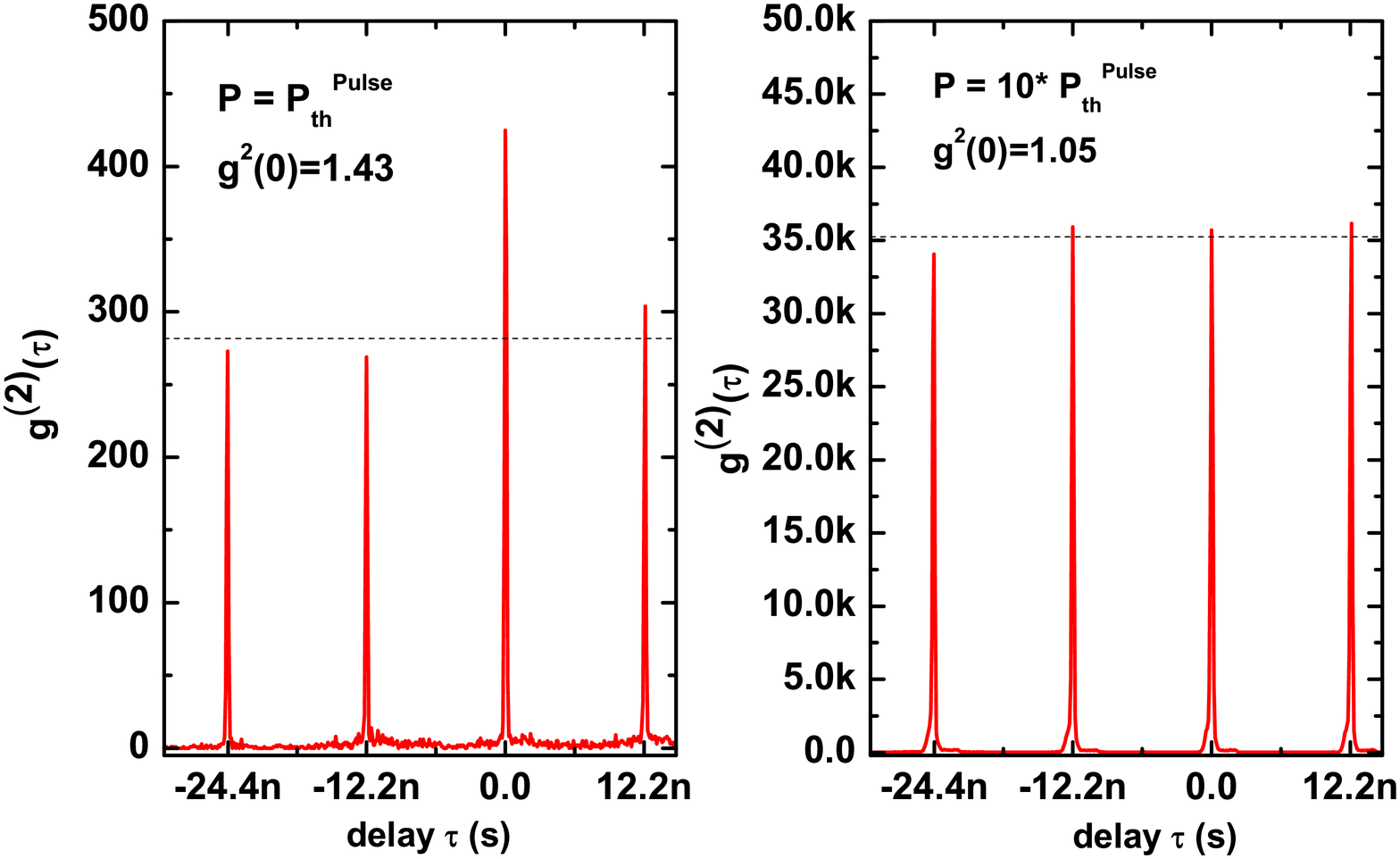}
   \end{tabular}
   \end{center}
   \caption[example]
   { \label{Fig3} Coincidence histogram as function of the delay between the two photons for two pump powers. (Left) at threshold, (Right) at ten times the threshold power}
   \end{figure}
   
Laser emission under pulsed excitation is validated by measuring the second order autocorrelation function $g^{(2)}(\tau)$ defined as
\begin{equation}
g^{(2)}(\tau) = \frac{<:I(t)I(t+\tau):>}{<I(t)>^2}
\end{equation}
where I(t) is the light intensity and $<::>$ corresponds to the normal order of the electric fields. For chaotic light $g^{(2)}(0)>1$ while for coherent light $g^{(2)}(0)$=1. Figure \ref{Fig3} depicts the histogram of the arrival times between two subsequent photons under pulsed excitation. As expected, distinct peaks separated by the pump laser repetition period is observed.  Two autocorrelation functions, at different excitation powers are shown. For a pump power at threshold (P=P$^{Pulse}_{th}$), the zero delay peak exhibits a 50\% increase on the integrated area, indicating as expected the emission of chaotic light. For an excitation power of P=10*P$^{Pulse}_{th}$, all peaks have the same integrated intensity, indicating the emission of coherent light. All peaks at $\tau$=iT $\{i \neq 0\}$, expect the peak at zero delay, are of equal integrated area $A_i=A_j \forall\{i,j\neq0\}$, since they correspond to uncorrelated coincidence events. One can therefore derive $g^{(2)}(0)$ by dividing the area of the zero delay $A_0$ by the mean value of $A_i \forall \{i\neq 0\}$.

 \begin{figure}[h!]
   \begin{center}
   \begin{tabular}{c}
   \includegraphics[width=8.3cm]{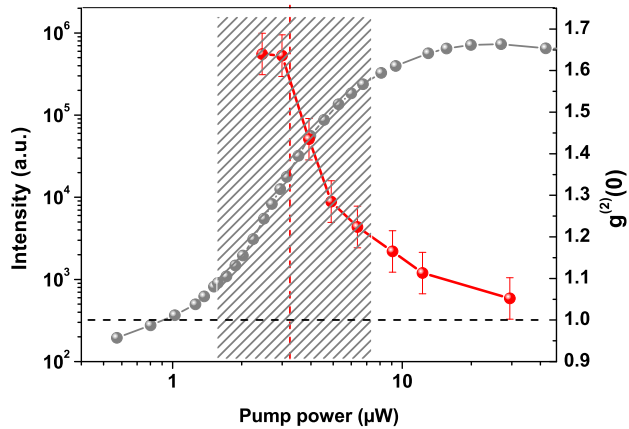}
   \end{tabular}
   \end{center}
   \caption[example]
   { \label{Fig4} Second order autocorrelation function $g^{(2)}(0)$ as function of the pump power. The shaded area indicates the transition region in the L-L curve}
   \end{figure}
   
Figure \ref{Fig4} represents the normalized autocorrelation function at zero delay (g$^{(2)}$(0)). A clear transition from chaotic $g^{(2)}(0)=1.64$ to coherent light is observed $g^{(2)}(0)\approx1$. The shaded area corresponds to the transition area in the L-L curve. It is noteworthy that although one would expect coherent emission at threshold, near the kink in the L-L curve at P=P$^{Pulse}_{th}$ or at least once the nonlinear increase ceases near P=2.68*P$^{Pulse}_{th}$. Coherent emission is obtained at a pump power of P$^{Pulse}_{coh}$=27.7 $\mu$W = 8.8*P$^{Pulse}_{th}$. Similar behavior has been observed in previous work on micropillar cavities \cite{Ulrich2007,Ates2008} or photonic crystal cavities \cite{Choi2007} based on InAs/GaAs quantum dots and operating at 4K. Extracting the value of $\beta$ using the phenomenological equation presented in \cite{Choi2007} requires fitting the g$^{(2)}(0)$ curve while having a precise knowledge of many experimental parameters, which are inaccessible in our setup, such as the coherence time of the quantum dots (T$_2$), or the total number of dipoles involved in the lasing action. By simply comparing the extend of the threshold region with all the previously published papers, we estimate a $\beta$ factor higher than $\beta>0.1$.

In summary we have demonstrated lasing from a double heterostructure cavity with InAsP/InP quantum dots as gain medium, at room temperature. The laser cavity is subjected to cQED effects and display high-$\beta$ values. Monomode lasing emission was achieved under both pulsed and continuous wave excitation, with a sideband mode rejection of more than 27 and 35 dB respectively. Onset of lasing was demonstrated by measuring the second order autocorrelation function at zero time delay. We demonstrate that coherent emission is achieved at a pump power six times greater than the threshold pump power defined as the kink in the Light-In Light-Out curve.

\end{document}